\documentstyle[12pt]{article} \textwidth 17cm \textheight 23 cm
\def\1{\mbox{I\hspace{-.15em}1}}

\def\b{\begin{equation}}
\def\e{\end{equation}}
\def\bee{\begin{enumerate}}
\def\eee{\end{enumerate}}

\title{Power Spectrum in Krein Space Quantization}
\author{M. Mohsenzadeh$^1$, S. Rouhani$^{1}$ and M.V. Takook$^{1}$\thanks{e-mail:
takook@razi.ac.ir}, }
\date{\today}

\begin{document}

\maketitle {\it \centerline{$^1$ Plasma Physics Research Center,
Islamic Azad University ,}  \centerline{\it P.O.BOX 14835-157,
Tehran, IRAN} }

\begin{abstract}

The power spectrum of scalar field and space-time metric
perturbations produced in the process of inflation of universe, have
been presented in this paper by an alternative approach to field
quantization namely, Krein space quantization \cite{gareta,ta2002}.
Auxiliary negative norm states, the modes of which do not interact
with the physical world, have been utilized in this method. Presence
of negative norm states play the role of an automatic
renormalization device for the theory.

\end{abstract}

\vspace{0.5cm} {\it Proposed PACS numbers}: 04.62.+v, 03.70+k,
11.10.Cd, 98.80.H \vspace{0.5cm}


\section{Introduction}

The minimally coupled scalar field in de Sitter space plays an
important role in the inflationary model as well as in the linear
quantum gravity \cite{gagata}. As proved by Allen \cite{al}, a
covariant quantization of minimally coupled scalar field cannot be
constructed by positive norm states alone. To cope with this problem
the Krein space quantization has been utilized \cite{gareta}.
Preserving the covariance principle and ignoring the positivity
condition, we have performed the field quantization in the Krein
space, a combined Hilbert and anti-Hilbert space \cite{mi}. This
process resembles Gupta-Bleuler quantization of the electrodynamic
equations in Minkowski space. It has been proven that the use of the
two sets of solutions (positive and negative norms states) is a
necessary feature for preservation of $(1)$ causality (locality),
$(2)$ covariance, and $(3)$ elimination of the infrared divergence
for the minimally coupled scalar field in de Sitter space
\cite{gareta,ta2001}. The ultraviolet divergence in the stress
tensor disappears as well, in other words the quantum free scalar
field in this method is automatically renormalized. The effect of
``unphysical'' states (negative norm states) appears in the above
theory as a natural renormalization mechanism.

Following above studies and through the same new approach, the
spectrum of density perturbation produced during inflation, has
been calculated in this paper. In the next section we briefly
recall definition of the spectrum of density fluctuation. Sections
$3$ is devoted to calculation of the spectrum of scalar field in
Krein space quantization. The spectrum of metric perturbation is
calculated in section $4$. Brief conclusion and outlook are given
in conclusion.

\section{Notation}

In the linear schema of evolution of universe, the density
perturbation is presented to be a function governed by simple
stochastic properties. For a Gaussian distribution these are
specified completely by the spectrum of density fluctuation. The
cosmologists are interested in the density contrast and the quantity
$ R $ which measures the associated spatial curvature perturbation.
Let $ f $ denote any one of these perturbations. Its most important
property is its spectrum, which is essentially the smoothed
modulus-squared of its Fourier coefficient. To be precise, the
spectrum may be defined as the quantity \cite{ja} \b
P_{f}\equiv(Lk/2\pi)^3
4\pi\langle|f_\textbf{\textbf{\textbf{k}}}|^2\rangle,\e where L is
the size of the box for the Fourier expansion, and the bracket
denotes the average over a small region of k-space. The
normalization is chosen to give a simple formula for the dispersion
(the root mean square vacuum fluctuation) of $ f $, which we shall
denote by $\ \sigma_{f}$. From the Fourier expansion one has
$\sigma_{f}^{2}=\sum\langle|f_{\textbf{\textbf{k}}}|^{2}\rangle $,
which lead to \cite{lily2000} \b \sigma_{f}^2\equiv\langle
f^2(\textbf{x})\rangle=\int_{0}^\infty P_{f}(k)\frac{dk}{k}.\e $
P_{f}$ is the spectrum, and the bracket now denoting the spatial
average.

We briefly recall the quantization of the massless scalar field
$\phi_p(x)$ in de Sitter space in the following coordinates \b
ds^{2}=g^{dS}_{\mu\nu}dx^\mu
dx^\nu=dt^{2}-a^2(t)d\textbf{x}^2=dt^{2}-e^{2Ht}d\textbf{x}^2,\e
where $a(t)$ is a scale factor and $H=\frac{\dot{a}}{a}$ is the
Hubble parameter. The scalar field operator $\phi_p(x)$ can be
represented in the form \b \phi_p(x)=(2\pi)^{-3/2}\int d^{3}k[c(k)
u_k(t)e^{i\textbf{k.x}}+c^{\dag}(k)u^*_k(t)e^{-i\textbf{k.x}}],\e
where $ u_k(t) $ satisfies the equation \cite{bomomo}
\b\ddot{u}_{k}+3H\dot{u}_{k}+\frac{k^{2}}{a^{2}}u_{k}=0. \e  The
exact solutions of the field equation are \cite{er,lin}\b
 u_{k,p}=\frac{e^{-ik\tau}}{a\sqrt{2k}}(1+\frac{i}{k\tau}),
\;\;u_{k,n}=\frac{e^{+ik\tau}}{a\sqrt{2k}}(1-\frac{i}{k\tau}), \e
where modes $u_{k,p}$ ($u_{k,n}$) represent positive (negative)
norm states and $\tau $ is conformal time $\frac{1}{aH}=-\tau.$
Then the quantity $ \langle \phi_p^{2}\rangle $ may be simply
expressed in terms of $ u_{k,p} $ of the above solution: \b
\langle\phi_p^{2}\rangle=\frac{1}{(2\pi)^3}\int
|u_{k,p}|^{2}d^{3}k=\frac{1}{(2\pi)^3}\int\frac{d^{3}k}{k}(\frac{1}{2a^{2}}+\frac{H^2}{2k^2}).
\e The first term is the usual contribution from vacuum
fluctuations in Minkowski space. This contribution can be
eliminated by renormalization. Accordingly the spectrum of the
scalar field, defined by $(2)$, after the renormalization of the
first term of the $(7)$, is given by \b
P_{\phi}(k)=\frac{H^{2}}{4\pi^{2}}. \e The metric perturbation
spectrum is also \cite{er} \b
P_{\Re}=\frac{H^{4}}{4\pi^{2}\dot{\phi}^{2}}.\e

\section{Spectrum of scalar field in Krein space}

In this section we calculate the spectrum of scalar field
perturbations in Krein space quantization thoroughly. First, we
briefly recall the Krein space quantization. In the previous paper
\cite{gareta,ta2002}, we present the free field operator in the
Krein space quantization. The field operator in Krein space is
build by joining two possible solutions of field equations,
positive and negative frequency states \b \phi(x)=\frac{1}{2}
[\phi_p(x)+\phi_n(x)],\e where
$$ \phi_p(x)=(2\pi)^{-3/2}\int d^{3}k [c(k)u_{k,p}(t)e^{i\textbf{k.x}}+c^{\dag}(
k)u_{k,p}^*(t)e^{-i\textbf{k.x}}],$$ $$
\phi_n(x)=(2\pi)^{-3/2}\int d^{3}k [b(
k)u_{k,n}(t)e^{-i\textbf{k.x}}+b^{\dag}(k)u_{k,n}^*(t)e^{i\textbf{k.x}}].$$
 $c(k)$ and $b(k)$ are two independent operators. Creation and
annihilation operators are constrained to obey the following
commutation rules \b [c( k),c^{\dag}(k')]=\delta(k-k') ,\;\; [b(
k),b^{\dag}( k')]=-\delta( k-k') .\e The other commutation
relations are all equal to zero. The vacuum state $\mid \Omega>$
is then defined by \b c^{\dag}(k)\mid \Omega>= \mid 1_{
k}>;\;\;c(k)\mid \Omega>=0, \e \b b^{\dag}( k)\mid \Omega>= \mid
\bar1_{k}>;\;\;b( k)\mid \Omega>=0, \e where $\mid 1_{k}> $ is
called a one particle state and $\mid \bar1_{k}>$ is called a one
``un-physical state''.

By imposing the physical interaction on the field operator, only
the positive norm states are affected. The negative modes do not
interact with the physical states or real physical world, thus
they can not be affected by the physical interaction as well. The
auxiliary negative norm states in our problem are similar to the
ghost states in the standard gauge QFT. In the gauge QFT, the
auxiliary negative norm states (ghost states), can neither
propagate in the physical world nor interact with physical states.
In the previous works, we have shown that presence of negative
norm states play the role of an automatic renormalization device
for certain problems \cite{gareta,ta2002,ta3,ta2,ta5,rota1}.

In the Minkowski space, the ultraviolet divergence of the vacuum
energy of the quantum field is removed by the normal ordering. In
the curved space-time, the standard renormalization of the
ultraviolet divergence of the vacuum energy is accomplished by
subtracting the local divergencies of Minkowski space (eq. (4.5)
in \cite{bida}),
$$<\Omega\mid:T_{\mu\nu}:\mid\Omega>=<\Omega\mid T_{\mu\nu}\mid\Omega>-<0\mid
T_{\mu\nu}\mid0>,$$ $\mid\Omega>$ is the vacuum state in curved
space and $\mid0>$ is vacuum state in Minkowski space. The mines
sign in the above equation can be interpreted as the negative norm
states which is added to the field operator. This interpretation
resembles Krein space quantization where the negative norm states
are considered as well. These negative norm states however, are
defined in Minkowski space alone and they are not the solution of
the wave equation in the curved space time. In other words this
renormalization vividly breaks the symmetry of the curved space.

There are other possibilities for removing the local divergencies
in the curved space-the so called non-standard renormalization
schemata-the local divergencies of curved space is removed by the
quantities defined in the very same curved space-time. In this
case, the mines sign can be interpreted as the negative norm
states which is added to the field operator and they are the
solutions of the wave equation in the curved space-time. This
scheme seams to be far more logical since the curved space
symmetry has been preserved after renormalization procedure. In
the previous or ''standard'' renormalization procedure the vacuum
is defined globally while the singularities are removed locally!

An alternative interpretation of above method is as follows. In
the curved space-time ($g_{\mu\nu}$), the negative norm states
could be constructed based upon two different perspectives, which
are defined with respect to the choice of the space-time
background field $(g^{BG}_{\mu\nu})$
$$ g_{\mu\nu}=g^{BG}_{\mu\nu}+h_{\mu\nu}=
 \eta_{\mu\nu}+h_{\mu\nu}^{Mi}=g^{dS}_{\mu\nu}+h_{\mu\nu}^{dS},$$
where $h_{\mu\nu}^{Mi}$ and $h_{\mu\nu}^{dS}$ are perturbation on
the minkowskian and de Sitter background respectively. In the first
perspective the background metric is the minkowskian
$g^{BG}_{\mu\nu}=\eta_{\mu\nu}$. In order to confirm the equivalence
of the results of the previous works with findings of this paper, we
have considered the negative norm states in the minkowskian space.
In this case, the physical result is similar to the renormalization
of the QFT in curved space-time with respect to the minkowskian
space-time \cite{bida}. In the second perspective the background
metric is the de Sitter space, $g^{BG}_{\mu\nu}=g^{dS}_{\mu\nu},$
and the physical result is similar to the non-standard
renormalization of the QFT in curved space-time with respect to the
de Sitter space-time. The negative norms are defined with respect to
the de Sitter background \cite{gareta}. These perspectives lead to
construction of two independent field operators.

\subsection{First perspective}

The negative norms are defined with respect to the minkowskian
background $\eta_{\mu\nu}$. The equations of the negative norm
states and their solutions are:
\b\ddot{v}_{k}+\frac{k^{2}}{a^{2}}v_{k}=0, \;\;\;
v_{k,n}=\frac{e^{+ik\tau}}{a\sqrt{2k}}.\e The scalar field
operator in such a perspective (through Krein quantization method)
can be written as: $$ \phi(x)=(2\pi)^{-3/2}\int d^3k \{[
c(k)u_{k,p}(t)e^{+i\textbf{k.x}}+c^{\dag}(k)u_{k,p}^*(t)e^{-i\textbf{k.x}}]$$
\b+[
b(k)v_{k,n}(t)e^{-i\textbf{k.x}}+b^{\dag}(k)v_{k,n}^*(t)e^{i\textbf{k.x}}]\}.\e

Then quantity $ \langle \phi^{2}\rangle $ may be simply expressed
in terms of above solutions: $$
\langle\phi^{2}\rangle=\frac{1}{(2\pi)^{3}}\int|u_{k,p}|^{2}d^{3}k-\frac{1}{(2\pi)^{3}}\int|v_{k,n}|^{2}d^{3}k
$$ \b =\frac{1}{(2\pi)^{3}}\int\frac{d^{3}k}{k}(\frac{1}{2a^{2}}
+\frac{H^{2}}{2k^{2}})-\frac{1}{(2\pi)^{3}}\int\frac{d^{3}k}{2ka^{2}}
=\frac{1}{(2\pi)^{3}}\int\frac{d^{3}k}{k}(\frac{H^{2}}{2k^{2}}) \e
Accordingly the spectrum, defined by (2), is given by \b
P_{\phi}=\frac{H^{2}}{4\pi^{2}}. \e This result is same as $(8)$.

\subsection{Second perspective}

The negative norm states are defined with respect to the curved
space-time background $g^{dS}_{\mu\nu}$. The scalar field operator
in this perspective (through Krein quantization method) could be
written as \cite{gareta}:
$$ \phi(x)=(2\pi)^{-3/2}\int d^3k \{[
c(k)u_{k,p}(t)e^{i\textbf{k.x}}+c^{\dag}(k)u_{k,p}^*(t)e^{-i\textbf{k.x}}]$$
\b+[
b(k)u_{k,n}(t)e^{-i\textbf{k.x}}+b^{\dag}(k)u_{k,n}^*(t)e^{i\textbf{k.x}}]\},\e
where $ u_{k,p}$ and $u_{k,n}$ are defined in $(6)$. The two-point
function for the above field operator \cite{ta2001}:
 \b  {\cal W}(x,x')=\frac{iH^2}{8\pi} \epsilon (t-t')[\delta(1-{\cal
 Z}(x,x'))-\theta ({\cal Z}(x,x')-1)], \e
 where
  $$ \epsilon (x^0-x'^0)=\left\{\begin{array}{clcr} 1&t>t'
 \\
  0&t=t'\\  -1&t<t'\\    \end{array} \right. ,$$
 and $\theta$ is the Heaviside step function. ${\cal
Z}(x,x')=\cosh H\sigma$ and $\sigma$ is the invariant geodesic
distance between the points $x$ and $x'$. It is clear that the
propagator is well defined and free of any divergencies (apart
from the delta function that does not appear in the present
calculation any way). Then quantity $ \langle \phi^{2}(x)\rangle $
may be simply expressed in terms of the time-ordered product
two-point function in coinciding points \cite{ta2} \b \langle
\phi^{2}(x)\rangle =\langle 0|T\phi(x) \phi(x)|0\rangle=iG_T(x -
x)=i\Re G_F(x,x). \e $G_F$ is usual Feynman propagator (eq. (9.52)
in \cite{bida}) \b \Re G_F(x,x) \approx \frac{-a_1(x)}{16
\pi}=\frac{H^2}{8\pi}.\e Using equations $(19)$ and $(21)$, we
obtain
 \b\langle\phi^{2}\rangle=\frac{iH^2}{8\pi}, \; \;\langle|\phi|^{2}
 \rangle=\frac{H^2}{8\pi}. \e Accordingly the spectrum, defined by $(2)$, easily is given by:
\b P_{\phi}=\frac{H}{4\pi^2} ke^{-\alpha k^{2}}, \e where $
\alpha=\frac{1}{\pi H^{2}}$. It is clear that the scale invariance
is broken here. This is similar to the previous works where the
scale invariance is broken due to consideration of perturbation of
the gravitational field \cite{kaka}.

It is important to note that in the second perspective a novel
cutoff (the exponent) appears which is related to the
characteristic dS scale $\alpha$. This significant difference
stems from the choice of dS background as opposed to the
minkowskian counterpart.

\section{Spectrum of space-time metric perturbation}

The linear theory of cosmological perturbations (namely the scalar
field and gravitational field perturbation considered here)
represent a cornerstone of modern cosmology and is used to describe
the formation and evolution of structures in the universe. The seeds
of these inhomogeneities were generated during inflation and
stretched over astrophysical because of a phase of rapid expansion
the universe. Study of inflation is special from the point of view
of perturbations. Any perturbation in the scalar field means a
perturbation to the energy-momentum tensor, $
\delta\phi\Longrightarrow \delta{T}_{\mu\nu}.$ A perturbation in the
energy-momentum tensor implies through the Einstein's equation of
motion, a perturbation of the metric \b
[\delta{R}_{\mu\nu}-\frac{1}{2}\delta(g_{\mu\nu}R)]=8\pi{G}\delta{T}_{\mu\nu}\Longrightarrow
\delta{g}_{\mu\nu}. \e Therefore we conclude that the perturbations
of the scalar field and that of the metric are tightly coupled to
each other and have to be studied together. The spectrum of the
metric perturbations is \cite{lily2000,itzu} \b
P^{1/2}_{\Re}=\frac{H}{\dot{\phi}}P^{1/2}_{\phi}. \e Therefore, the
power spectrum for metric perturbations at horizon scale in the
first perspective is given by \b
P_{\Re}=\frac{H^{4}}{4\pi^{2}\dot{\phi}^{2}},\e and in the second
perspective is
 \b
P_{\Re}\cong\frac{H^{3}}{4 \pi^2\dot{\phi}^{2}}ke^{-\alpha k^{2}}.
\e Similar to the previous case, a dS-induced cutoff (the
exponent) appears which is connected with the characteristic dS
scale $\alpha$. The emergence of a cutoff in this result is
completely new. The physical reason for the different behavior of
the spectra in these two perspectives is the breaking of de Sitter
invariance in the first perspective \cite{al} and its preservation
in the second \cite{gareta}. One of the physical implications of
such a ''cutoff'' is the calculation of the CMBR anisotropy, which
will be considered in the forthcoming paper.

\section{Conclusion}

The negative frequency solutions of the field equations are needed
for the covariant quantization in the minimally coupled scalar
fields in de Sitter space. Contrary to Minkowski space, the
elimination of de Sitter negative norms in this case breaks the de
Sitter invariance. In other words, to restore the de Sitter
invariance, one needs to take into account the negative norm
states {\it i.e.} the Krein space  quantization. This provides a
natural tool for renormalization of the theory \cite{gareta}. In
the present paper, the spectrum of the scalar field perturbations,
has been calculated through the Krein space quantization. It is
found that the theory is automatically renormalized. The negative
norm states could be constructed based upon two different
perspectives. In the first case negative norm states could be
defined in flat space-time similar to renormalization of the QFT
in curved space-time \cite{bida}. The standard result is obtained
in this case, eqs. $(8)$ and $(9)$. In the second perspective, the
negative norm states are defined with respect to the curved
space-time. In this case the scale invariance is broken. The scale
invariance is also broken due to consideration of perturbation of
the interaction field \cite{kaka}.

In the second perspective a dS-induced cutoff appears which is
connected with the characteristic dS scale $\alpha$- in the first
scheme this cutoff is missing. In other words the global
properties of the Minkowski and the dS backgrounds do result in
this specific departure. The different behavior of the spectra in
these two perspectives is due to the breaking of de Sitter
invariance in the first perspective and its preservation in the
second. In the forthcoming paper the effect of this spectrum on
CMBR anisotropy will be discussed in detail. \vspace{0.5cm}

\noindent {\bf{Acknowlegements}}:  The authors would like to thank
A. Sojasi and E. Yusofi for their interest in this work.

\end{document}